\documentclass[12pt,a4paper]{article}

\pdfoutput=1

\usepackage[margin=3.5cm]{geometry}
\usepackage{amsmath,amssymb,graphicx}
\usepackage[dvipsnames]{xcolor}
\usepackage{hyperref}
\usepackage{multicol}
\usepackage{soul}
\usepackage{slashed}
\usepackage{appendix}
\usepackage{changepage}
\usepackage{comment}
\usepackage{authblk}
\usepackage{scalefnt}
\usepackage{setspace}
\usepackage{subcaption}

\allowdisplaybreaks

\numberwithin{equation}{section}
\begin{document}
	

\title{\bf Worldline Images for Yang-Mills Theory within Boundaries}

\author{Santiago Christiansen Murguizur, Lucas Manzo, Pablo Pisani\vspace{10pt}}

\affil{ Departamento de Física, Facultad de Ciencias Exactas
	(UNLP)\vspace{-10pt}}

\affil{ Instituto de Física La Plata
	(UNLP/CONICET) Argentina}

\date{}

\maketitle

\thispagestyle{empty}

\begin{abstract}
	In this article we develop a worldline technique based on the method of images to study the effective action associated to Yang-Mills theories on manifolds with boundaries. We consider the possibility of having either relative or absolute boundary conditions, which are particular types of mixed boundary conditions. Both vector fields and ghost fields are taken into account in this analysis. As a check of our construction, we compute the first three Seeley-DeWitt coefficients of the heat kernel asymptotics. Finally, we employ our technique to calculate the rate of gluon production due to a chromoelectric field background in the presence of a boundary.
\end{abstract}

\vfill

\noindent\rule{5cm}{.25mm}

\vspace*{-3mm}

\noindent{\footnotesize santiago.christiansen@fisica.unlp.edu.ar\\[-4mm]
lucasmanzo@fisica.unlp.edu.ar\\[-4mm]
pisani@fisica.unlp.edu.ar}


\pagebreak

\tableofcontents

\vspace{7mm}
\hrule
\vspace{2mm}

\section{Introduction}

In quantum field theory the effective action of any specific model --given by its fundamental fields and interactions-- can be constructed using Feynman diagrams up to the desired perturbative order. Nevertheless, under non-dynamical external conditions --such as interactions with background fields, with the spacetime metric or with boundaries-- the functional methods \cite{Vassilevich:2003xt} might result more convenient. In this framework the 1-loop effective action is computed through the functional determinant of the differential operator whose spectrum describes the quantum fluctuations. Among the several techniques to evaluate functional determinants, the versatile worldline formalism \cite{Schubert:2001he,Bastianelli:2006rx,Corradini:2015tik} has proved adaptable to different types of fields on curved spacetimes: scalars \cite{Bastianelli:1991be,Bastianelli:2002fv}, spinors \cite{Bastianelli:2002qw,Bastianelli:2003bg} (see also \cite{Bastianelli:1992ct,McKeon:1993np}), and vectors \cite{Bastianelli:2005vk,Bastianelli:2005uy} (where, more generally, antisymmetric tensor fields of arbitrary rank are considered) have been studied with worldline techniques. Also higher-spin fields on conformally flat spaces \cite{Bastianelli:2008nm,Corradini:2010ia,Bastianelli:2012bn} and quantum gravity calculations have been addressed from this perspective \cite{Bastianelli:2013tsa,Bastianelli:2019xhi}. In the last couple of decades an increasingly large number of contributions laid the groundwork for the further development of this formalism providing the tools for its extension to other scenarios (see the concise reviews \cite{Edwards:2019eby,Schubert:2023bed} and references therein). More recently, in the last couple of years there has been different improvements in the approach to scalar fields \cite{Bonezzi:2025iza,Kim:2025hwi}, vectors \cite{Bonezzi:2024emt,Bastianelli:2025xne} (see \cite{Dai:2008bh}) and gravitons \cite{Bonezzi:2020jjq,Bastianelli:2023oyz,Fecit:2024jcv}. Also recently axial couplings to many different background fields have been worked out \cite{Bastianelli:2024vkp}.

In parallel to this line of research, many other applications have been developed within the worldline framework, such as nonperturbative phenomena \cite{Affleck:1981bma,Dunne:2005sx,Dunne:2006st,Gordon:2014aba,Gordon:2016ldj,Fecit:2025kqb,Bastianelli:2025khx,Choi:2025jis,Copinger:2025ovz} and numerical methods \cite{Gies:2001zp,Gies:2001tj,Schmidt:2002mt,Gies:2003cv,Gies:2005sb,Edwards:2017ndv,Ahumada:2025poa}. The formalism has also been found to be particularly suited to study noncommutative geometries \cite{Bonezzi:2012vr,Vinas:2014exa,Ahmadiniaz:2015qaa,Ahmadiniaz:2018olx}. In a recent work the worldline techniques have also been adapted to the so called covariant fracton gauge theories \cite{Fecit:2025eet}. Currently, the connection between worldline methods and classical scattering of black {holes} has sparked an intense research on its application to gravity \cite{Goldberger:2004jt,Porto:2007qi,Kalin:2020mvi,Mogull:2020sak,Jakobsen:2021smu,Jakobsen:2022psy,Kim:2024grz,Kim:2024svw}.

The implementation of the worldline formalism to quantum fields within boundaries has been pursued by some of the authors of the present article and collaborators. In \cite{Bastianelli:2006hq,Bastianelli:2007jr,Bastianelli:2008vh,Bastianelli:2009mw} the case of a scalar field in the presence of a flat boundary under different boundary conditions and from different perspectives has been solved. Semitransparent interfaces have been considered in \cite{Vinas:2010ix,Ahmadiniaz:2022bwy}.

The case of a curved boundary required additional strategies; the first application was developed in \cite{Corradini:2019nbb} for a scalar field in a specific geometry, {namely, the $D$-dimensional ball}. In \cite{Manzo:2024gto} one of the authors of the present article extended these techniques to {spinor fields under bag boundary conditions on the $D$-dimensional manifold $\mathbb{R}^{D-1}\times \mathbb{R}^+$ ($x^D=0$ denotes the boundary) equipped with a metric such that $g_{\alpha D}=0$, for all $\alpha\neq D$. The present article completes this program by considering nonabelian vector fields on these geometries. The choice of these geometric assumptions is motivated, a priori, by the collar neighborhood structure of a manifold in the vicinity of its boundary. However, our technical procedure --based on the image method-- forces us to impose this structure as a global property of the manifold. Still, many scenarios are encompassed in this context as, for example, the ``deformation of a single flat boundary'' or, in more precise terms: the flat submanifold of $\mathbb{R}^D$ defined by $x^D\geq f(x^\alpha)$, for any single-valued function $f:\mathbb{R}^{D-1}\to\mathbb{R}$ of the remaining coordinates $x^\alpha$. Moreover, one would expect that the computation of a local quantity under our assumptions gives some insight into the behavior of this same quantity near the boundary in more general cases as well (for instance, the boundary contribution to the local rate of gluon production computed in section \ref{sec cbf}).}

We analyze two types of gauge invariant boundary conditions --known as {\it absolute} and {\it relative}-- and give a worldline representation of the 1-loop effective action. To illustrate our result, we compute the first few Seeley-DeWitt coefficients of the heat kernel. 

As a second application, we calculate the rate of gluon production by an external chromodynamic electric field $E$ parallel to a boundary. We obtain the well known bulk contribution proportional to $|E|^2$ plus a new term localized at distance  $\sim|E|^{-1/2}$ from the boundary. In our worldline representation these contributions correspond to the actions of worldline instantons of two types: those confined in the bulk of the manifold and those that suffer reflections at the boundary.

{The procedure we follow in this article is known as top-down in the worldline context, in the sense that we start from a specific model in QFT --in our case, pure Yang-Mills with boundaries-- and we look for a worldline representation of the operator of quantum fluctuations of the field. In contrast, bottom-up procedures depart from the quantization of a system in 0+1 dimensions and determine to which field theory it corresponds by analyzing the constraints imposed by the symmetries of this point-particle (see e.g.\ \cite{Bastianelli:2005vk}).}

In the worldline formalism the effective action for a field $\Phi(x)$ on a spacetime $M$ is written in terms of the path integral over the closed trajectories $x^\mu(t)$ of an auxiliary point particle. In this way, physical quantities in a quantum field theory are expressible in terms of elements of a theory in first quantization. For the case of a quantum field confined by boundaries $\partial M$, the trajectories of the auxiliary particle are also confined within such boundaries. At this point one realizes that the standard techniques of perturbative calculation of path integrals are not directly applicable to trajectories lying in spaces with boundaries. In particular, one must design a practical procedure to integrate over functions $x^\mu(t)$ which take values on a target space which is bounded. Moreover, a wave function on a space with boundaries satisfies a specific boundary condition --usually, for a given type of field, there is a family of infinitely many admissible boundary conditions. In canonical first quantization the specific boundary condition is imposed on the solutions of the equation of motion but how does one impose the boundary condition in the path integral formulation? In other words, in the context of path integral quantization, how does one restrict the set of trajectories in $M$ or modify the corresponding measure $\mathcal Dx^\mu(t)$ in order to obtain a transition amplitude that satisfies a certain boundary condition? In our previous articles we have proposed some general techniques aimed at these questions. Their particular answers for the case of a nonabelian vector field --in particular, regarding the specific boundary conditions-- is the main result of the present work.

The presentation of the article is as follows. In section \ref{review} we provide a rough but self-contained review of the computation of the 1-loop effective action of pure Yang-Mills fields in the presence of a boundary. Boundary conditions for both gauge and ghost fields are derived. In section \ref{WL} we put forward a worldline representation of the effective action of Yang-Mills theory with boundaries in terms of the transition amplitudes in first quantization of a particle. To illustrate an application of our representation, we compute in section \ref{sec ht} the first few Seeley-DeWitt coefficients, which are related to the small proper-time expansion of the transition amplitude. As a second application, related rather to the semiclassical approximation of the path integral than to a {perturbative} expansion, we compute in section \ref{sec cbf} the rate of gluon production due to a background chromoelectric field. Finally, in section \ref{sec conclu} we summarize our findings, draw some conclusions and propose further research.

\section{Yang-Mills theory with a boundary}\label{review}

We begin with a succinct review of the 1-loop analysis of quantum fluctuations in pure Yang-Mills theory on Euclidean manifolds with boundaries. This section follows the presentation in \cite{Vassilevich:2003xt} (see also \cite{Avramidi:1997hy}).

Throughout this article we restrict ourselves to a base Riemannian manifold $M$ that admits global coordinates $x^\mu$ (with $\mu=1,2,\dots,D$) on the $D$-dimensional half-space $\mathbb{R}^{D-1}\times \mathbb{R}^+$. We use the first Greek letters $\alpha, \beta, \gamma,\ldots=1,2,\ldots,D-1$ to denote coordinates on $\partial M$, located at $x^D=0$.  {We assume} the metric $g_{\mu\nu}$ satisfies $g_{\alpha D}=0$, so the induced metric on the boundary is $h_{\alpha\beta}=g_{\alpha\beta}$; the normal inward unit vector is then $n^\mu=\sqrt{g^{DD}} \delta^\mu_D$.

We introduce a gauge field $A_\mu(x)=A_\mu^I(x)\,T^I$, where $T^I$ ($I=1,2,...,N$) are the generators of some Lie algebra $[T^I,T^J]=if^{IJK}\,T^K$, chosen, as is usual, such that $f^{IJK}$ is totally antisymmetric and ${\rm tr}\,(T^I T^J)=\frac12\,\delta^{IJ}$. The Yang-Mills action of this field is
\begin{align}
	S[A]=\frac12\int_M d^Dx\,\sqrt{g}\ {\rm tr}\,(F^{\mu\nu}F_{\mu\nu})\,,
\end{align}
where $F_{\mu\nu}=-i[D_\mu,D_\nu]$, in terms of the covariant derivative $D_\mu=\partial_\mu+iA_\mu$.

Following the background field method we make the shift $A_\mu\to A_\mu+a_\mu$, separating quantum fluctuations $a_\mu$ from a fixed background $A_\mu$. Accordingly, the field tensor splits into a background field tensor, and both linear and quadratic perturbations,
\begin{align}\label{chafmn}
	F_{\mu\nu}\to F_{\mu\nu}+\nabla_\mu a_\nu-\nabla_\nu a_\mu+i[a_\mu,a_\nu]\,.
\end{align}
Note we are here using the full covariant derivative
\begin{align}
	\nabla_\mu a^I_\nu=\partial_\mu a^I_\nu-\Gamma^\rho_{\mu\nu}a^I_\rho-f^{IJK}A^J_\mu a^K_\nu\,,
\end{align}
which --together with the gauge connection with respect to the background field $A_\mu$-- includes the Levi-Civita connection as well. This replacement of $D_\mu$ by $\nabla_\mu$ is possible because the antisymmetric combination of the second and third terms in \eqref{chafmn} cancels the contribution of the Christoffel symbols.

The 1-loop effects of quantum fluctuations are encoded in the terms of $S[A+a]$ which are quadratic in $a_\mu$; these are readily obtained from the square of \eqref{chafmn},
\begin{align}
	S^{(2)}[A,a]&=\int_M d^Dx\,\sqrt{g}
	\ {\rm tr}\left(
	\nabla_\mu a_\nu \nabla^\mu a^\nu
	-\nabla_\mu a_\nu \nabla^\nu a^\mu
	+iF_{\mu\nu}[a^\mu,a^\nu]
	\right)
	\,.
\end{align}
To write this expression in terms of the quadratic form of an elliptic operator one proceeds as follows. In the first term one simply integrates by parts. In the second term one also integrates by parts $\nabla_\mu$ but then interchanges the order of the covariant derivatives in $\nabla_\mu\nabla_\nu$ to finally integrate by parts back again, this time $\nabla_\nu$. The consequent commutator $[\nabla_\mu,\nabla_\nu]$ gives a term proportional to $R_{\mu\nu}$ plus another term, proportional to $F_{\mu\nu}$, which adds to the fourth term. One thus obtains
\begin{align}\label{scuad}
	S^{(2)}[A,a]&=\int_M d^Dx\,\sqrt{g}
	\ {\rm tr}\left\{-a_\mu\nabla^2a^\mu
	-(\nabla_\mu a^\mu)^2
	+a^\mu R_{\mu\nu}a^\nu-2ia^\mu[F_{\mu\nu},a^\nu]\right\}
	+\mbox{}\nonumber\\[2mm]
	&\mbox{}+\int_{\partial M} d^{D-1}x\,\sqrt{h}
	\ n^\mu\,{\rm tr}\left\{
	-a^{\nu}\left(\nabla_\mu a_\nu-\nabla_\nu a_\mu\right)
	-a_\mu\,\nabla_\nu a^\nu\right\}
	\,.
\end{align}
The three integrations by parts produce their respective boundary terms. We now choose the gauge $\nabla_\mu a^\mu=0$, which introduces a term proportional to $(\nabla a)^2$ into the action with an arbitrary coefficient. The Feynman-'t Hooft gauge for this coefficient is the choice that cancels both contributions proportional to $(\nabla a)^2$ and gives the following elliptic operator for the dynamics of the quantum fluctuations:
\begin{align}\label{H}
	{\mathcal D_{\nu}\mbox{}^{\mu\,IJ}=-\delta^\mu_\nu\,\delta^{IJ}\,\nabla^2
	+R_{\nu}^\mu\,\delta^{IJ}+2f^{IJK}F_{\nu}\mbox{}^{\mu\,K}\,.}
\end{align}
In addition, the dynamics of the ghost fields associated to our gauge choice is dictated by the operator {$\mathcal B^{IJ}=-\delta^{IJ}\,\nabla^2$}. The spectrum of the quantum fluctuations of the gauge and ghost fields, $a_\mu(x)$ and $c(x)$, are the eigenvalues of $\mathcal D_{\mu\nu}^{IJ}$ and $\mathcal B^{IJ}$.

To turn $\mathcal D$ and $\mathcal B$ into symmetric operators, we must choose boundary conditions --on the normal component $a_D(x)$, the tangential components $a_\alpha(x)$, and the ghost fields $c(x)$-- for which the boundary terms in \eqref{scuad} vanish. These terms can be written as
\begin{align}\label{operator}
	\int_{\partial M} d^{D-1}x\,\sqrt{h}
	\ n^D\,{\rm tr}\left\{
	-a^{\alpha}\left(D_D a_\alpha-D_\alpha a_D\right)
	-a_D\,\nabla_\nu a^\nu\right\}\,.
\end{align}
There are many possibilities of getting rid of the boundary terms; we will analyze two of them. One can either take Dirichlet conditions on the tangential components, $a_\alpha=0$, and then impose
\begin{align}
	0&=\nabla_\nu a^\nu=\partial_D a^D+\Gamma_{DD}^D a^D+i[A_D,a^D]+\Gamma_{\alpha D}^\alpha a^D\nonumber\\[2mm]
	&=\sqrt{g_{DD}}\left(n^\mu\nabla_\mu-L\right) a^D\,,
\end{align}
where $L=L^\alpha_\alpha=-\sqrt{g^{DD}}\,\Gamma_{\alpha D}^\alpha$ is the trace of the second fundamental form $L_{\alpha\beta}=n_\mu \Gamma^\mu_{\alpha\beta}$. This set of Dirichlet conditions for $a_\alpha^I$ and Robin conditions for $a_D^I$ is known as relative (or magnetic) boundary conditions. In 4-dimensional Minkowski spacetime these reproduce the well-known behavior at a ``perfect conductor'', namely, normal chromoelectric and tangent chromomagnetic fields at $x^D=0$.

Alternatively, one can impose Dirichlet conditions on the normal component, $a_D=0$. This must be complemented with
\begin{align}
	0&=D_D a_\alpha-D_\alpha a_D=\partial_D a_\alpha\nonumber\\[2mm]
	&=\sqrt{g_{DD}}\left(n^\mu\nabla_\mu a_\alpha-L^\beta_\alpha a_\beta\right)\,.
\end{align}
Note that, for consistency, we have assumed $A_D=0$. This set is called absolute (or electric) boundary conditions and, in 4-dimensional Minkowski spacetime, they lead to a tangential chromoelectric field and a normal chromomagnetic field at the boundary.

On the basis of gauge invariance we must finally determine --for both relative and absolute conditions-- the appropriate boundary conditions on the ghost fields $c(x)$. In other words, we analyze conditions on the parameter $\varepsilon(x)$ of gauge transformations $a_\mu(x) \to a_\mu(x)+D_\mu\varepsilon(x)$ which preserve the boundary conditions on $a_\mu$. Ghost fields inherit the boundary conditions of the gauge parameter.

Let us consider first absolute boundary conditions. The boundary condition $n^\mu\nabla_\mu \varepsilon=0$ certainly preserves $a_D=0$ {(using, once more, $A_D=0$)}. Since absolute boundary conditions are also assumed on the background field, the parameter satisfies $\partial_D\varepsilon=0$ at the boundary. Gauge invariance of $\partial_Da_\alpha=0$ thus arises from using $\partial_DA_\alpha=0$ in $\partial_DD_\alpha\varepsilon=i[\partial_DA_\alpha,\varepsilon]=0$. The condition $n^\mu\nabla_\mu c=0$ on the ghost fields therefore preserves gauge invariance of absolute boundary conditions.

As regards relative boundary conditions, $\varepsilon=0$ (for which $\partial_\alpha\varepsilon=0$) clearly preserves $a_\alpha=0$. The remaining condition $n^\mu\nabla_\mu a_D=L\,a_D$ is maintained due to $n^\mu\nabla_\mu D_D\varepsilon=n^D\nabla_D \nabla_D\varepsilon=(\sqrt{g_{DD}}\,\nabla^2+L)\varepsilon$. If $\varepsilon(x)$ is an eigenfunction of the Laplacian then $\nabla^2\varepsilon=0$ at the boundary and the condition for $a_D$ is also preserved. As a consequence, Dirichlet conditions on the ghost fields preserve relative boundary conditions under gauge transformations.

Let us summarize the two sets of boundary conditions that will be considered in this article:
\begin{align}
	&{\rm Relative\ b.c.:}
	\qquad a_\alpha=\left(n^\mu\nabla_\mu-L\right) {a_D}=c=0\,,\\[2mm]
	&{\rm Absolute\ b.c.:}
	\qquad a_D=n^\mu\nabla_\mu a_\alpha-L^\beta_\alpha a_\beta=n^\mu\nabla_\mu c=0\,.
\end{align}
Under these conditions both $\mathcal D$ and $\mathcal B$ are symmetric operators. We have now laid down the setting to compute the quantum effective action $\Gamma[A]$, which is defined through
\begin{align}\label{effact}
	e^{-\frac{1}{\hbar}\Gamma[A]}=\int\mathcal Da\ e^{-\frac{1}{\hbar}S[A+a]}
	\,e^{\frac{1}{\hbar}\int d^Dx\,J^\mu a_\mu}\,.
\end{align}
Here $J^\mu(x)$ is the external current that generates the background field $A_\mu(x)$,
\begin{align}
	\frac{\delta \Gamma[A]}{\delta A_\mu(x)}=J^\mu(x)\,.
\end{align}
By expanding $S[A+a]$ in expression \eqref{effact} around the background field $A_\mu(x)$ one obtains that the leading quantum effects on $\Gamma[A]$ arise from quadratic terms in $a_\mu(x)$,
\begin{align}
	\Gamma[A]=S[A]-\hbar\log\,\int\mathcal Da\ e^{-\frac12\,S^{(2)}[A,a]}+O(\hbar^2)
\end{align}
where $S^{(2)}[A,a]$ is given in \eqref{scuad}. As described at the beginning of this section, upon gauge fixing the quadratic integral one gets that the 1-loop contributions to the effective action are given by the functional determinants of the operators $\mathcal D$ and $\mathcal B$ (we omit higher order terms in $\hbar$),
\begin{align}\label{eff act 2}
	\Gamma[A]&=S[A]+\frac \hbar2\,\log{{\rm Det}\,\mathcal D}-\hbar\,\log{{\rm Det}\,\mathcal B}\nonumber\\[2mm]
	&=S[A]-\frac \hbar2\int_0^\infty \frac{dT}{T}
	\left({\rm Tr}\,e^{-T\mathcal D}-2\,{\rm Tr}\,e^{-T\mathcal B}\right)\,.
\end{align}
In the second line we have used Schwinger proper-time representation for the functional determinants. Traces and determinants must be computed with the appropriate boundary conditions on the domains of $\mathcal D$ and $\mathcal B$.

In the next section we construct a path-integral representation of the kernel of the operators $e^{-T\mathcal D}$ and $e^{-T\mathcal B}$ for relative and absolute boundary conditions.

\section{Worldline representation}\label{WL}

In this section we set forth a method for applying worldline techniques to the scenario described in the previous section: a gauge field $A^I_\mu(x)$ ($I=1,...,N$) on a manifold $M$ with metric $g_{\mu\nu}$ parametrized by coordinates on the half-space $x=(x^\alpha,x^D)$, where $x^\alpha\in\mathbb{R}^{D-1}$ and $x^D\in\mathbb{R}^+$. On the boundary $\partial M$, at $x^D=0$, the field satisfies either absolute or relative boundary conditions. The main subject of this section is to present a device to impose these boundary conditions by means of an appropriate set of auxiliary trajectories.

Both conditions can be written as
\begin{align}\label{genbc}
	{\Pi^- a(x)=\Pi^+(n^\mu\partial_\mu-S)\,a(x)=0\qquad {\rm at\ }x\in\partial M\,.}
\end{align}
The operator $\Pi^-$ projects onto the tangential (normal) components for relative (absolute) boundary conditions; $\Pi^+=1-\Pi^-$ is its complementary projection. We define the operator $\chi=\Pi^+-\Pi^-$, which will be mostly important in the sequel. Explicitly, the components of the matrix $S=\Pi^+ S=S\Pi^+$ are given by $S_{\nu}^{\;\mu}\mbox{}^{IJ}=[(n^\rho\Gamma_{\rho D}^D+L)\,\delta^{IJ}-f^{IJK}n^\rho A_\rho^K]\,\delta_\nu^D \delta^\mu_D$ for relative conditions and $S_\nu^{\;\mu}\mbox{}^{IJ}=0$ for absolute boundary conditions. {Operators $\Pi^\pm$ and $\chi$ are diagonal in the gauge indices.}

Our procedure is based on appropriately extending operators on $M$ to a twofold version $\tilde M$ parametrized by the whole $\mathbb{R}^D$; $\tilde M$ can be thought of as two copies of $M$ glued together along the boundary $\partial M$. Path integral representations on the space $\tilde M$ --which has no boundaries-- are already known.

In fact, let us consider a $D\,N\times D\,N$ differential operator $\mathcal O(\hat x,\hat p)$ on sections of the vector bundle defined over a space that, like $\tilde M$, is parametrized by the whole $\mathbb R^D$. The Hermitian momentum operator is defined as {\cite{DeWitt:1952js}}
\begin{align}
	\hat p_\mu=-ig^{-\frac14}\partial_\mu\,g^\frac14\,,
\end{align}
and we normalize the position and momentum bases as
\begin{align}
	\langle p|p’\rangle=\delta^{(D)}(p-p’)\;\;\;\;\text{and}\;\;\;\; \langle x|x’\rangle=g^{-1/2}\delta^{(D)}(x-x’)\,.
\end{align}
The heat kernel of $\mathcal O(\hat x,\hat p)$ admits the following phase-space integral representation \cite{Bastianelli:2006rx},
\begin{align}\label{intrep}
	\langle x'|e^{-T\mathcal O(\hat x,\hat p)}|x\rangle
	=\left[g(x)g(x')\right]^{-\frac14}
	\int \mathcal Dx(t)\,\mathcal Dp(t)
	\ \mathcal P
	\,e^{-\int_0^T dt\left(\mathcal O_W(x,p)-ip_\mu \dot x^\mu\right)}\,.
\end{align}
The integration trajectories $x(t)$ satisfy $x(0)=x$ and $x(T)=x'$; no such restrictions hold on $p(t)$. The symbol $\mathcal P$ represents the path ordering.

Ordering ambiguities are solved through the Weyl ordering $\mathcal O_W(\hat x,\hat p)$ of the operator $\mathcal O(\hat x,\hat p)$. In this notation, $\mathcal O_W(\hat x,\hat p)=\mathcal O(\hat x,\hat p)$ (as operators) but in the former $\hat x$ and $\hat p$ are commuted into a symmetric expression. For example, if $\mathcal O(\hat x,\hat p)=\hat x\hat p$ then $\mathcal O_W(\hat x,\hat p)=\frac12\hat x\hat p+\frac12\hat p\hat x+\frac12[\hat x,\hat p]$. In other words, the operator must be cast into its Weyl form before the replacement $\hat x\to x(t)$ and $\hat p\to p(t)$ in \eqref{intrep} is performed. Therefore, certain counterterms might appear as a result of the required commutations. For the simple example $\mathcal O=\hat x\hat p$ one would get in the path integral $\mathcal O_W(x,p)=x(t)p(t)+\frac 12\,i$.

We now turn to $M$ and propose the following expression for the heat kernel of the operator $\mathcal D$ on gauge fields $a_\mu^I(x)\in \mathbb R^{D\times N}$ under boundary conditions of the type \eqref{genbc}:
\begin{align}\label{pirep}
	\langle x'|e^{-T\mathcal D}|x\rangle_M
	=\langle  x'|e^{-T\mathcal D_S}|x\rangle
	+\chi\, \langle \tilde x'|e^{-T\mathcal D_S}|x\rangle\,.
\end{align}
Here $x,x'\in M$. Amplitudes on the r.h.s.\ are computed on $\tilde M$ using \eqref{intrep}. The symbol $\sim$ above any point ${x}$ represents its reflection with respect to the boundary; that is, if  $x=(x^1,...,x^{D-1},x^D)$ then $\Tilde{x}=(x^1,...,x^{D-1},-x^D)$. The operator $\mathcal D_S$ on the r.h.s.\ contains a Dirac delta at the boundary,
\begin{align}\label{deltaS}
	\mathcal D_S=\tilde{\mathcal D} +2\sqrt{g^{DD}}\,S\,\delta(x^D)\,,
\end{align}
where $\tilde {\mathcal D}$ is the extension of $\mathcal D$ to $\tilde M$ defined as 
\begin{align}\label{extension}
	\tilde{\mathcal D}\,a(x)=\chi \mathcal{D}\chi\,a(\tilde x)
\end{align}
for $x^D<0$. It is important to remark that $\chi$ is originally defined at $\partial M$ so there is an implicit extension of the projectors from $\partial M$ to the whole $\tilde M$ such that they still satisfy $(\Pi^\pm)^2=\Pi^\pm$ and $\Pi^++\Pi^-=1$. We will assume this extension to be smooth and even with respect to the boundary. With some abuse in notation, we will also use $\Pi^\pm$ (as well as $\chi$) to denote their extensions.

Before proving \eqref{pirep} we should discuss the extension of the operator in a more concrete fashion. From \eqref{H} we note that $\mathcal D$ can be written as the scalar Laplacian plus a first order differential operator 
\begin{align}\label{partial deriv}
	\mathcal D=-\frac{1}{\sqrt{g}}\,\partial_\mu(\sqrt{g}g^{\mu\nu}\,\partial_\nu)
	+2\omega^\mu\partial_\mu+C\,,
\end{align}
with
\begin{equation}\label{omega}
	\left(\omega^\mu\right)^{\rho\,IJ}_\sigma=\delta^{IJ}g^{\mu\omega}\Gamma_{\omega\sigma}^\rho-\delta^\rho_\sigma f^{IJK}A^{\mu K}
\end{equation}
and
\begin{equation}
	{C_\rho\mbox{}^{\sigma\,IJ}=\left(\partial_\mu\omega^\mu+\omega^\mu\partial_\mu\text{log}\sqrt{g}
	-\omega^\mu\omega_\mu\right)_\rho\mbox{}^{\sigma\,IJ}+\delta^{IJ}R_\rho^{\sigma}+2f^{IJK}F_\rho\mbox{}^{\sigma\,K}\,.}
\end{equation}
The extension \eqref{extension} can then be written as 
\begin{align}\label{extended operator}
	\tilde{\mathcal D}=-\frac{1}{\sqrt{\tilde g}}\,\partial_\mu(\sqrt{\tilde g}\tilde{g}^{\mu\nu}\,\partial_\nu)
	+2\tilde{\omega}^\mu\partial_\mu+\tilde C\,,
\end{align}
where $\tilde g_{\mu\nu}$ is the symmetric extension of the metric, $\tilde g_{\mu\nu}(\tilde x)=\tilde g_{\mu\nu}(x)$ (see figure \ref{fig:metric4}), whereas the extensions of $C(x)$ and $\omega^\mu(x)$ to the whole $\tilde M$ are defined as
\begin{align}
	\tilde C(x)&=\chi C(\tilde x)\chi+\chi \omega^\alpha(\tilde x)\partial_\alpha\chi
	-\chi \omega^D(\tilde x)\partial_D\chi-\chi\frac{1}{\sqrt{g}}\partial_\mu(\sqrt{g}\partial^\mu\chi)\,,\label{Ctilde}\\[2mm]
	\tilde \omega^\alpha(x)&=\chi \omega^\alpha(\tilde x)\chi-2\chi\partial^\alpha\chi\,,\label{omegaalpha}\\[2mm]
	\tilde \omega^D(x)&=-\chi \omega^D(\tilde x)\chi-2\chi\partial^D\chi\,,\label{omegaD}
\end{align}
for $x^D<0$.
As one can see, for an arbitrary extension of $\chi$ the previous expressions are rather cumbersome. However, if $\chi$ is extended as a constant matrix they turn out to be easy to picture. {If we take} $\chi=\pm\text{diag}(- 1,\ldots,-1, 1)$ --the upper (lower) sign corresponds to relative (absolute) boundary conditions-- on the whole $\tilde M$ then $C^D_D$, $C^\alpha_\beta$ and $\omega^D$ are symmetrically extended, whereas $C^D_\alpha$, $C^\alpha_D$ and $\omega^\alpha$ become antisymmetric with respect to the interchange $x\leftrightarrow\tilde x$.   

We must now show that our ansatz \eqref{pirep} satisfies the heat equation\footnote{The subindex in $\mathcal D_{x'}$ indicates that the operator acts on the first argument of the kernel.}
\begin{align}\label{heat equation}
	\mathcal{D}_{x’}\langle x'|e^{-T\mathcal D}|x\rangle_M+\partial_T \langle x'|e^{-T\mathcal D}|x\rangle_M=0\,,
\end{align} 
the initial condition
\begin{align}\label{initial}
	\langle x'|e^{-T\mathcal D}|x\rangle_M\,\bigg|_{T=0}=
	{g(x)^{-\frac12}}\ \delta^{(D)}(x-x’)
\end{align}
and the boundary conditions \eqref{genbc} at $x’\in\partial M$.

Equation \eqref{heat equation} follows readily from the application of $\partial_T$ to \eqref{pirep} and the use of $\chi(\mathcal D_S)_{\tilde x'}=(\mathcal D_S)_{x'}\chi$. The proof of \eqref{initial} is also straightforward.

We check the Dirichlet boundary condition in \eqref{genbc} explicitly,
\begin{align}\label{proof dir}
		\Pi^-\langle x'|e^{-T\mathcal D}|x\rangle_M
		&=\Pi^-\langle  x'|e^{-T\mathcal D_S}|x\rangle 
		+\Pi^-\chi \langle \tilde x'|e^{-T\mathcal D_S}|x\rangle\nonumber\\[2mm]
		&=\Pi^-\langle  x'|e^{-T\mathcal D_S}|x\rangle 
		-\Pi^- \langle \tilde x'|e^{-T\mathcal D_S}|x\rangle\,.
\end{align}
The coefficients in $\mathcal{D}$ are assumed to be smooth functions on $M$ so those in $\tilde{\mathcal{D}}$ could have {at most finite jumps} at $\partial M$ due to the extension to $\tilde M$. As a consequence, $\langle  x'|e^{-T\mathcal D_S}|x\rangle$ must be continuous at $x'^D=0$ and therefore the r.h.s.\ of \eqref{proof dir} vanishes at $x'=\tilde x'$.

We must finally check that our ansatz \eqref{pirep} satisfies the Robin boundary condition in \eqref{genbc}. To this end we consider the heat equation satisfied by the r.h.s.\ of \eqref{pirep} but allowing $x,x’$ to be any pair of points in $\tilde M$,
{
\begin{equation}
	\left(\partial_T+\tilde{\mathcal{D}}_{x’}+2\sqrt{g^{DD}}\,S\,\delta(x’^D)\right)
	\left(\langle  x'|e^{-T\mathcal D_S}|x\rangle
	+\chi \langle \tilde x'|e^{-T\mathcal D_S}|x\rangle\right)=0\,.
\end{equation}
}
We now multiply by $-\tilde g_{DD}(x’)$ and integrate in $x’^D$ on a small interval $(-\varepsilon,\varepsilon)$ around $x'^D=0$. Because of the continuity of the heat kernels, as $\varepsilon\to 0$ we do only get contributions from the delta function and from $\partial_D'^2$ in $\tilde{\mathcal{D}}_{x'}$,
\begin{align}
	&\left( \partial'_D\big|_{x'^D=0^+}
	-\partial'_D\big|_{x'^D=0^-}
	-2\sqrt{g_{DD}}\ S\,\big |_{x’^D=0}\ \right)\times\mbox{}\nonumber\\[2mm]
	&\mbox{}\times\left(\langle  x'|e^{-T\mathcal D_S}|x\rangle
	+\chi \langle \tilde x'|e^{-T\mathcal D_S}|x\rangle\right)=0\,.
\end{align}
{Since the reflection of the r.h.s.\ of \eqref{pirep} with respect to $x'^D=0$ is implemented by $\chi$ (assumed to be symmetrically extended) then the difference between the lateral derivatives at $x'^D=0$ produces a factor $1+\chi=2\Pi^+$ and the previous expression can be cast into the form}
\begin{align}
	\left(\Pi^+ \partial'_D\big|_{x'^D=0^+}
	-\sqrt{g_{DD}}\ S\,\big |_{x’^D=0}\ \right)\left(\langle  x'|e^{-T\mathcal D_S}|x\rangle
	+\chi \langle \tilde x'|e^{-T\mathcal D_S}|x\rangle\right)=0\,.
\end{align}
Multiplying by $\sqrt{g^{DD}}$ and recalling that $S=\Pi^+S$ we finally obtain
\begin{equation}
	\Pi^+(n^\mu\partial'_\mu-S)\langle x’|e^{-T\mathcal D}|x\rangle_M\bigg |_{x’\in \partial M}=0\,,
\end{equation}
which completes our proof.

Following \eqref{intrep}, we can now use \eqref{pirep} to write down a path integral representation of the transition amplitude with boundary,
\begin{align}\label{pirep+intrep}
	&\langle x'|e^{-T\mathcal D}|x\rangle_M=\left[g(x){g(x')}\right]^{{-\frac14}}
	\int \mathcal Dx(t)\,\mathcal Dp(t)
	\ \mathcal P
	\,e^{-\int_0^T dt\left\{(\mathcal{D}_S)_W(x,p)-ip_\mu \dot x^\mu\right\}}+\mbox{}\nonumber\\
	&\mbox{}+\left[g(x){\tilde g(\tilde x’)}\right]^{-\frac14}\chi
	\int \mathcal Dx(t)\,\mathcal Dp(t)
	\ \mathcal P
	\,e^{-\int_0^T dt\left\{(\mathcal{D}_S)_W(x,p)-ip_\mu \dot x^\mu\right\}}\,.
\end{align}
{Trajectories satisfy $x(0)=x$, $x(T)=x'$ in the first integral, and $x(0)=x$, $x(T)=\tilde x'$ in the second one. There are no boundary conditions on $p(t)$.}

To compute ${(\mathcal{D}_{S})}_W$ we first consider the Weyl ordering of $\tilde{\mathcal{D}}$ (see \eqref{extended operator}),
\begin{equation}\label{weyl ordered extended operator}
	\tilde{\mathcal D}_W(\hat x,\hat p)=(\tilde g^{\mu\nu}(\hat x)\hat p_\mu\hat p_\nu )_S+\Delta H_e[\tilde g(\hat x)]+2i(\tilde\omega^\mu(\hat x)\hat p_\mu)_S+\Delta H_v(\hat x)\,,
\end{equation}
where the subscript $S$ indicates {symmetrization} with respect to $\hat x$ and $\hat p$. Neither $\Delta H_e$ nor $\Delta H_v$ contain $\hat p$. The former comes from Weyl ordering the first term in \eqref{extended operator}. {Replacing by classical variables, i.e.\ $\hat x\to x$,} these counterterms read
\begin{equation}
	\Delta H_e[\tilde g(x)]=\frac{1}{4}\left(-\tilde R+\tilde g^{\mu\nu}\tilde \Gamma_{\mu\rho}^\sigma\tilde \Gamma_{\nu\sigma}^\rho\right)\,,
\end{equation}
in terms of the curvature and connection of $\tilde g_{\mu\nu}$, the metric on $\tilde M$. This expression is already known from the fluctuation operator of a scalar field \cite{Bastianelli:2006rx}.

On the other hand the term
\begin{align}
	\Delta H_v(x)=\tilde C-\partial_\mu\tilde \omega^\mu-\tilde  \omega^\mu \partial_\mu \text{log}\sqrt{\tilde g}
\end{align}
is specific to the gauge theory (see \eqref{Ctilde}-\eqref{omegaD}). It contains many boundary terms originated in the commutator of taking derivatives and reflecting with respect to the boundary. To use $\mathcal{D}_{S}$ in the representation \eqref{pirep+intrep} we must still add the delta function term in \eqref{deltaS}. For simplicity, we define
\begin{align}\label{DeltaHS}
	\Delta H_S(x)=\Delta H_v+2\sqrt{g^{DD}}\,S\,\delta(x^D)\,.
\end{align}

Note that the analysis of this section can be easily extended to the ghost operator $\mathcal B$ after appropriate replacements. In fact, $\mathcal B$ takes the form \eqref{partial deriv} upon
\begin{align}
	&\left({\omega}^\mu\right)^{IJ}\to\left(\omega_{\text{gh}}^\mu\right)^{IJ}=-f^{IJK}A^{K\mu}\,,\\
	&C\to C_{\text{gh}}=
	\partial_\mu\omega^\mu_{\text{gh}}+\omega^\mu_{\text{gh}}\partial_\mu\text{log}\sqrt{g}-\omega^\mu_{\text{gh}}{\omega_{\text{gh}}}_\mu\,.
\end{align}
Moreover, the ghost boundary condition can also be written as \eqref{genbc} using $S_{gh}=0$ and $\Pi_{gh}^-=1$ ($\Pi^+_{gh}=0$ and $\chi_{gh}=-1$) for relative boundary conditions or $\Pi^-_{gh}=0$ ($\Pi^+_{gh}=1$ and $\chi_{gh}=1$) for absolute boundary conditions.

In the next sections we will use the representation \eqref{pirep+intrep} to compute some quantities of physical relevance in problems with boundaries.

\section{Heat trace expansion}\label{sec ht}

As an application of \eqref{pirep+intrep}, in this section we compute the first few Seeley-DeWitt coefficients $a_n$, up to $n=2$, which describe the small $T$ asymptotic expansion of the trace
\begin{align}\label{ht-trace}
	\text{Tr}\,e^{-T \mathcal{D}}&=\int_M d^Dx\,\sqrt{g(x)}\ \text{tr}\,\langle x| e^{-T\mathcal D}|x\rangle_M
	\sim T^{-D/2}\ \sum_{n=0}^\infty\ a_n\,T^{n/2}\,.
\end{align}  
The lower-case trace sums over both Lorentz and color indices of the gauge field. As can be seen from the proper-time regularization of the 1-loop effective action (see \eqref{eff act 2}) the first Seeley-DeWitt coefficients describe the UV behavior of the theory. To analyze the Yang-Mills case we will consider both the gauge and the ghost contributions.

Before computing the integrals in \eqref{pirep+intrep} it is convenient to perform the rescaling $t\to T\tau$, with $0<\tau<1$, together with the following shifts:
\begin{align}
	x^\mu(\tau)&\rightarrow x^\mu+\Delta x^\mu\,\tau+\sqrt{T}\,h^\mu(\tau)\,,\label{x}\\[2mm]
	p_\mu(\tau)&\rightarrow\frac{p_\mu(\tau)}{\sqrt{T}}+\frac{i}{2T}\,g_{\mu\nu}(x)\Delta x^\nu\,,\label{p}
\end{align} 
{where $x^\mu$ and $x^\mu+\Delta x^\mu$ denote the initial and endpoint of the trajectory $x^\mu(\tau)$, respectively}. The new integration variables $h(\tau)$, $p(\tau)$ are dimensionless and, as will be clear shortly, can be considered as $O(T^0)$ for small $T$. Note that $h(\tau)$ satisfies homogeneous Dirichlet conditions, $h(0)=h(1)=0$.

Inserting these shifts into \eqref{intrep} we obtain for the operator $\mathcal D_{S}$
\begin{align}\label{expr mean vaue}
		\langle x’|e^{-T\mathcal D_S}|x\rangle=
		\frac{[g(x)]^{\frac14}[\tilde g(x’)]^{-\frac14}}{(4\pi T)^{D/2}}e^{-\frac{1}{4T}g_{\mu\nu}(x)\Delta x^\mu\Delta x^\nu}\left\langle\mathcal{P} e^{-\int_0^1d\tau\, H_{\text{int}}(h(\tau),p(\tau))}\right\rangle\,,
\end{align}
where
\begin{align}
	H_{\text{int}}(h,p)&=\left(\tilde g^{\mu\nu}(x+\Delta x\, \tau+\sqrt{T}h(\tau))-g^{\mu\nu}(x)\right)\times\nonumber\\[2mm]
	&\times\left(p_\mu(\tau)+\frac{i}{2\sqrt{T}}\,g_{\mu\rho}(x)\Delta x^\rho\right)
	\left(p_\nu(\tau)+\frac{i}{2\sqrt{T}}\,g_{\nu\sigma}(x)\Delta x^\sigma\right)+\nonumber\\[2mm]
	&+T\Delta H_e[\tilde g(x+\Delta x \,\tau+\sqrt{T}h(\tau))]+\nonumber\\[2mm]
	&+2i\,\sqrt{T}\,\tilde \omega^\mu(x+\Delta x\, \tau+\sqrt{T}h(\tau))\left(p_\mu(\tau)+\frac{i}{2\sqrt{T}}g_{\mu\nu}(x)\Delta x^\nu\right)+\nonumber\\[2mm]
	&+T\Delta  H_S(x+\Delta x\,\tau+\sqrt{T}h(\tau))\,.
\end{align}
The mean value $\langle\ldots\rangle$ used in \eqref{expr mean vaue} is, more generally, defined as
\begin{equation}
	\langle f(h,p)\rangle=\frac{(4\pi T)^{D/2}}{\sqrt{g(x)}}\int\mathcal{D}h\mathcal{D}p\;
	e^{-\int_0^1d\tau\left\{ g^{\mu\nu}(x)p_\mu(\tau)p_\nu(\tau)-ip_\mu(\tau)\dot h^\mu(\tau)\right\}}
	f(h,p)\,.
\end{equation}
The appropriate normalization of the path integral has been determined through
\begin{align}
	\langle 1\rangle=
	\frac{(4\pi T)^{D/2}}{\sqrt{g(x)}}\,\langle 0| e^{-T\,g^{\mu\nu}(x)\hat p_\mu\hat p_\nu}|0\rangle
	=1\,.
\end{align}
Note that since $g^{\mu\nu}(x)$ is evaluated at a fixed point, it is constant and simply represents the mass of a freely moving particle.

It is convenient to compute the generating functional
\begin{align}\label{gen funct}
	&Z[k,j]=\left\langle e^{i\int_0^1d\tau\left(k^\mu(\tau)p_\mu(\tau)+j_\mu(\tau)h^\mu(\tau)\right)}\right\rangle\nonumber\\[2mm]
	&=e^{-\frac{1}{2}\int_0^1d\tau d\tau’\left(\frac{1}{2}g_{\mu\nu}(x)k^\mu(\tau)k^\nu(\tau’)+G(\tau,\tau’)g^{\mu\nu}(x)j_{\mu}(\tau)j_\nu(\tau’)
		+iG'(\tau,\tau’)k^\mu(\tau)j_\mu(\tau’)\right)}\,,
\end{align}
where
\begin{align}
	G(\tau,\tau’)&=-|\tau-\tau’|-2\tau\tau’+\tau+\tau'\\[2mm]
	{G'}(\tau,\tau’)&=-\epsilon(\tau-\tau')-2\tau'+1\,,
\end{align}
with $\epsilon(\cdot)$ the sign function. From expression \eqref{gen funct} one easily reads the two-point functions
\begin{align}
	\langle p_\mu(\tau)p_\nu(\tau’)\rangle&=\frac{1}{2}g_{\mu\nu}(x)\,,\label{two-pointpp}\\[2mm]
	\langle h^\mu(\tau)h^\nu(\tau’)\rangle&=g^{\mu\nu}(x)G(\tau,\tau’)\,,\label{two-pointhh}\\[2mm]
	\langle p_\nu(\tau)h^\mu(\tau’)\rangle&=\delta^\mu_\nu\,\frac{i}{2}\,G'(\tau,\tau’)\,.\label{two-pointph}
\end{align}

We will now compute the trace \eqref{ht-trace} by computing \eqref{pirep} using \eqref{expr mean vaue} order by order in $T$.

\subsection{Direct term}

\begin{figure}[t!]
    \centering
    \begin{minipage}{.8\linewidth}
    \begin{subfigure}[t]{.50\linewidth}
        \caption{}
        \includegraphics[width=.8\linewidth]{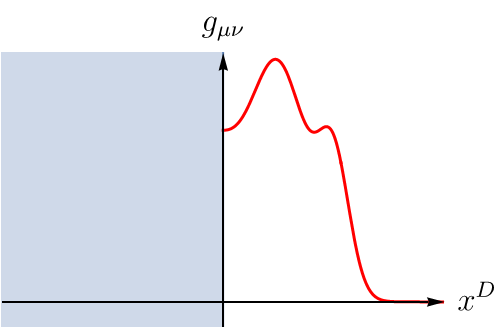}
        \label{fig:metric1}
    \end{subfigure}
    \begin{subfigure}[t]{.50\linewidth}
        \caption{}
        \includegraphics[width=.8\linewidth]{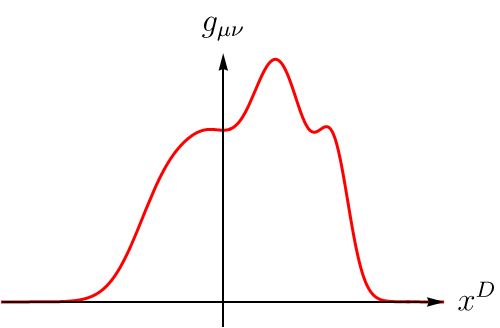}
        \label{fig:metric2}
    \end{subfigure}
    \begin{subfigure}[t]{.50\linewidth}
        \caption{}
        \includegraphics[width=.8\linewidth]{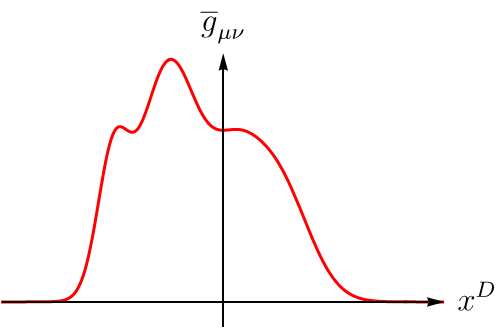}
        \label{fig:metric3}
    \end{subfigure}
    \begin{subfigure}[t]{.50\linewidth}
        \caption{}
        \includegraphics[width=.8\linewidth]{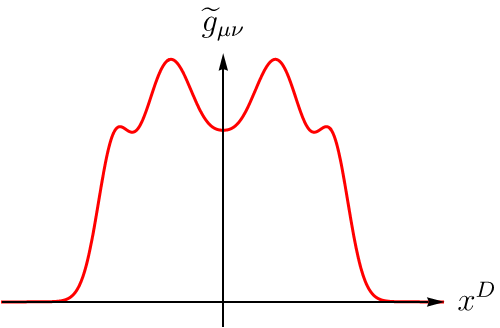}
        \label{fig:metric4}
    \end{subfigure}
    \caption{Visual representation of the different metrics defined in the present article, as a function of the normal coordinate $x^D$. Figure \ref{fig:metric1} is an arbitrary metric component $g_{\mu\nu}$ in $M$, while figure \ref{fig:metric2} is its analytic extension to $\tilde M$ (for simplicity we use the same symbol $g_{\mu\nu}$). Then $\bar g_{\mu\nu}$, depicted in figure \ref{fig:metric3}, is its reflection with respect to the boundary. Finally $\tilde g_{\mu\nu}$, depicted in figure \ref{fig:metric4}, is the symmetric extension of $g_{\mu\nu}$.}
    \label{fig:metrics}
    \end{minipage}\hspace{10mm}
\end{figure}

The two terms in \eqref{pirep} --which we call {\it direct} and {\it indirect} terms-- will be treated separately; in this section we will analyze the contribution of the direct term, for which $\Delta x=0$,
\begin{align}
	\mathbb{D}&=\int_M d^Dx\, \sqrt{g(x)}\ \text{tr}\langle x| e^{-T{\mathcal D}_S}|x\rangle\nonumber\\
	 &=\frac{1}{(4\pi T)^{D/2}}\int_M d^Dx\,\sqrt{g(x)}\ \text{tr}\,\left\langle\mathcal{P} e^{-\int_0^1d\tau H_{\text{int}}(h(\tau),p(\tau))}\right\rangle\,.
\end{align} 
Here $H_{\text{int}}$ reduces to 
\begin{align}
	H_{\text{int}}(h,p)=&\left(\tilde g^{\mu\nu}(x+\sqrt{T}h)-g^{\mu\nu}(x)\right)p_\mu p_\nu+
	T\Delta H_e[\tilde g(x+\sqrt{T}h)]+\nonumber\\
	&+2i\,\sqrt{T}\,\tilde \omega^\mu(x+\sqrt{T}h)\,p_\mu+T\Delta H_S(x+\sqrt{T}h)\,.
\end{align}
This expression indicates that $H_{\text{int}}=O(\sqrt{T})$ so, to obtain all coefficients up to $a_2$, we expand
\begin{align}\label{direct expanded}
	\mathbb{D}=&\frac{1}{(4\pi T)^{D/2}}\int_M d^Dx\sqrt{g(x)}\ \text{tr}\,
	\bigg(1-\int_0^1d\tau \left\langle H_{\text{int}}(h(\tau),p(\tau))\right\rangle+\nonumber\\[2mm]
	&+\frac{1}{2}\int_0^1d\tau d\tau’ \left\langle H_{\text{int}}(h(\tau),p(\tau))H_{\text{int}}(h(\tau’),p(\tau’))
	\right\rangle+\ldots\bigg)\,.
\end{align}
For smooth coefficients one would Taylor expand $H_{\text{int}}$. However --due to our extension to $\tilde M$-- $H_{\text{int}}$ has finite discontinuities as well as delta-type singularities (stemming from first derivatives as well as from the term containing $S$ in \eqref{DeltaHS}) at $x^D=0$. To appropriately deal with them note that in extending $g_{\mu\nu}$ as $\tilde g_{\mu\nu}$, odd powers of the normal coordinate in $g_{\mu\nu}$ change sign at $x^D=0$. For all $x\in\tilde M$ we now use $g_{\mu\nu}(x)$ to denote the analytic extension of the metric, that is, with no sign change in the odd power of $x^D$. With this notation, we write
\begin{equation}
	\tilde g^{\mu\nu}(x)=g^{\mu\nu}(x)-\Theta(-x^D)\,\delta g^{\mu\nu}(x)\,,
\end{equation}
where $\Theta(x)$ is the Heaviside step function and {$\delta g^{\mu\nu}(x)=g^{\mu\nu}(x)- g^{\mu\nu}(\tilde x)$}. In this way $g^{\mu\nu}(x)$ and $\delta g^{\mu\nu}(x)$ are smooth functions on the whole $\tilde M$ and singularities are isolated into $\Theta(x)$ and its derivatives. In particular, for $\Delta H_e$ we obtain
\begin{align}
	\Delta H_e[\tilde g(x)]&=\Delta H_e[ g(x)]-\Theta(-x^D)(\Delta H_e[ g(x)]-\Delta H_e[\bar g(x)])
	+\mbox{}\nonumber\\[2mm]
	&-\sqrt{g^{DD}(x)}\,\delta(x^D)\,L\,.
\end{align}
Here $\bar g_{\mu\nu}(x)=g_{\mu\nu}(\tilde x)$ so $\bar g_{\mu\nu}(x)=\tilde g_{\mu\nu}(x)$ for $x^D<0$; in this way, we make explicit that the difference $\Delta H_e[ g(x)]-\Delta H_e[\bar g(x)]$ is smooth in $x$. To visualize the difference between $g_{\mu\nu}$, $\bar g_{\mu\nu}$ and $\tilde g_{\mu\nu}$, figure \ref{fig:metrics} contains an example for an arbitrary metric component.

Similarly, interpreting $\omega^\mu(x)$ as the analytic extension to $\tilde M$ and writing
\begin{equation}
	\tilde \omega^\mu(x)=\omega^\mu(x)-\Theta(-x^D)\,\delta\omega^\mu(x)\,,
\end{equation}
with
\begin{align}
	\delta\omega^\alpha(x)&=\omega^\alpha(x)-\chi\omega^\alpha(\tilde x)\chi+2\chi\partial^\alpha\chi\,,\\[2mm]
	\delta\omega^D(x)&=\omega^D(x)+\chi\omega^D(\tilde x)\chi+2\chi\partial^D\chi\,,
\end{align}
we isolate the step-like and delta-like discontinuities in $\Delta  H_S$,
\begin{equation}
	\Delta  H_S(x)=\Delta H_v(x)-\Theta(-x^D)\Delta H_\theta(x)+\delta(x^D)
	\left({-\delta\omega^D(x)}+2\sqrt{g^{DD}}S\right).
\end{equation}
We do not write down the expression for the coefficient $\Delta H_\theta$ explicitly because we will not use it in the sequel.

All in all, we can separate the contributions to $H_{int}$ as (i) ``bulk terms'', i.e., smooth contributions, (ii) terms containing a step function, located at $x^D<0$, and (iii) terms proportional to a delta function, supported at the boundary:
\begin{align}\label{3terms}
		H_{\text{int}}(h,p)=&H_b(h,p)- H_\theta(h,p)\,\Theta(-x^D-\sqrt{T}h^D)+H_\delta(h)\,\delta(x^D+\sqrt{T}h^D)\,.
\end{align}
Here
\begin{align}
		H_b(h,p)&=\left( g^{\mu\nu}(x+\sqrt{T}h)-g^{\mu\nu}(x)\right)p_\mu p_\nu+
		T\Delta H_e[ g(x+\sqrt{T}h)]+\mbox{}\nonumber\\[2mm]
		&+2i\,\sqrt{T}\, \omega^\mu(x+\sqrt{T}h)\,p_\mu+T\Delta H_v(x+\sqrt{T}h)\,,\label{hb}
\end{align}
\begin{align}
		H_\theta(h,p)&= \delta g^{\mu\nu}(x+\sqrt{T}h)\ p_\mu p_\nu+\mbox{}\nonumber\\[2mm]
		&+T(\Delta H_e[ g(x+\sqrt{T}h)]-\Delta H_e[ \bar g(x+\sqrt{T}h)])+\mbox{}\nonumber\\[2mm]
		&+2i\,\sqrt{T}\, \delta\omega^\mu(x+\sqrt{T}h(\tau))\,p_\mu+T\Delta H_\theta(x+\sqrt{T}h(\tau))\,,\label{ht}\\[2mm]
		H_\delta(h)&=T\bigg(-\sqrt{g^{DD}(x+\sqrt{T}h)}\ L(x+\sqrt{T}h){-\delta\omega^D(x+\sqrt{T}h)}+\mbox{}\nonumber\\[2mm]
		&+2\sqrt{g^{DD}(x+\sqrt{T}h)}\ S(x+\sqrt{T}h)\bigg)\label{hd}\,.
\end{align}
These functions can be Taylor expanded safely. We can now compute the different contributions to \eqref{direct expanded} corresponding to each of the three terms in \eqref{3terms} separately.

We begin with
\begin{align}
	\mathbb{D}_b=&\frac{1}{(4\pi T)^{D/2}}\int_M d^Dx\sqrt{g(x)}\;\text{tr}\,\bigg(1-\int_0^1d\tau \left\langle H_{b}(h(\tau),p(\tau))\right\rangle+\nonumber\\
	&+\frac{1}{2}\int_0^1d\tau d\tau’ \left\langle H_{b}(h(\tau),p(\tau))H_{b}(h(\tau’),p(\tau’))\right\rangle\bigg)\,.
\end{align}
The calculation goes as follows: we expand $H_b$ in powers of $\sqrt{T}$, we compute expectations values using \eqref{two-pointpp}-\eqref{two-pointph}, and we integrate in $\tau$ and $\tau’$. After some grouping work we get
\begin{align}\label{Db}
			\mathbb{D}_b=\frac{1}{(4\pi T)^{D/2}}\int_M d^Dx\sqrt{g}\;\text{tr}\,\bigg(1+T\bigg[ \frac{R}{6}-C
			+\partial_\mu \omega^\mu
			+\omega^\mu\partial_\mu\text{log}\sqrt{ g}-\omega^\mu\omega_\mu\bigg]\bigg)\,.
\end{align}
Next, we compute
\begin{align}
	\mathbb{D}_\delta=-\frac{1}{(4\pi T)^{D/2}}\int_M d^Dx\sqrt{g}\int_0^1d\tau \;\text{tr}\left\langle H_{\delta}(h(\tau))\delta(x^D+\sqrt{T}h^D(\tau))\right\rangle\,.
\end{align}
It is convenient to rescale $x^D\rightarrow \sqrt{T}x^D$ and then expand in $T$ (for this contribution the leading order is sufficient),
\begin{align}\label{Ddelta}
		\mathbb{D}_\delta&=-\frac{T}{(4\pi T)^{D/2}}\int_M d^Dx\,\sqrt{g(x^\alpha,0)}\int_0^1d\tau \;\left\langle\delta(x^D+h^D(\tau))\right\rangle\times\nonumber\\[2mm]
		&\times\text{tr}\left(\sqrt{g^{DD}(x^\alpha,0)}\left(2S(x^\alpha)-L(x^\alpha)\right)
		-\delta\omega^D(x^\alpha,0)\right)\,.
\end{align}
To compute the mean value of the delta function we write
\begin{align}
	\left\langle\delta(x^D+h^D(\tau))\right\rangle=\frac{1}{2\pi}\int_{-\infty}^\infty dk\, e^{-ikx^D}\left\langle e^{-ik h^D(\tau)}\right\rangle\,,
\end{align}
where
\begin{align}
	\left\langle e^{-ik h^D(\tau)}\right\rangle
	=Z[0,-ik\delta^D_\mu\delta(\tau'-\tau)]
	=e^{-\frac{k^2}{2}g^{DD}(x^\alpha,0)G(\tau,\tau)}\,.
\end{align}
One then obtains
\begin{align}
	\left\langle\delta(x^D+h^D(\tau))\right\rangle
	=\frac{e^{-\frac{(x^D)^2}{2g^{DD}(x^\alpha,0)G(\tau,\tau)}}}{\sqrt{2\pi g^{DD}(x^\alpha,0)G(\tau,\tau)}}\,.
\end{align}
Plugging this into \eqref{Ddelta}, expanding the remaining $T$ dependence, and integrating in $x^D$ and $\tau$ one gets
\begin{align}
	\begin{aligned}
		\mathbb{D}_\delta=&-\frac{T}{2(4\pi T)^{D/2}}\int_{\partial M} d^{D-1}x\,\sqrt{h}\,\text{tr}\left(2S-\sqrt{g_{DD}}\,\delta\omega^D-L\right)\,.
	\end{aligned}
\end{align}
From the definition \eqref{omegaD} and using $\text{tr}(\chi\,\partial_D\chi)=-\text{tr}(\partial_D\chi\,\chi)=0$ we obtain $\text{tr}(\delta\omega^D)=2\,\text{tr}(\omega^D)$. We then write
\begin{align}\label{Dd}
		\mathbb{D}_\delta=-\frac{T}{2(4\pi T)^{D/2}}\int_{\partial M} d^{D-1}x\sqrt{h}\,\text{tr}\left(2S-2\sqrt{g_{DD}}\,\omega^D-L\right)\,.
\end{align}

We finally compute
\begin{align}
	\mathbb{D}_\theta=&\frac{1}{(4\pi T)^{D/2}}\int_M d^Dx\sqrt{g}\int_0^1d\tau \;\text{tr}\left\langle H_{\theta}(h,p)\,\Theta(-x^D-\sqrt{T}h^D)\right\rangle\,.
\end{align}
We follow the same steps as for $\mathbb{D}_\delta$, namely, we rescale $x^D\rightarrow \sqrt{T}x^D$ and expand in $T$. For the mean value of the step function we use, as before, the Fourier decomposition
\begin{align}
	\Theta(-x)=\int_{-\infty}^\infty\frac{dk}{2\pi i}\ \frac{e^{-ikx}}{k-i0}\,.
\end{align}
The resulting mean values are obtained from the generating function as
\begin{align}
		\bigg\langle &h^{\mu_1}(\tau_1)...h^{\mu_n}(\tau_n)\,p_{\nu_{1}}(\eta_{1})\ldots
		p_{\nu_{m}}(\eta_{m})\,e^{-ik h^D(\tau)}\bigg\rangle=\nonumber\\[2mm]
		&=(-i)^{n+m}\prod_{(i,j)=(1,1)}^{(n,m)}\frac{\delta\;}{\delta  k^{\nu_j}(\eta_j)}\frac{\delta\;}{\delta  j_{\mu_i}(\tau_i)}Z[k,j]\bigg|_{k^\mu(\tau’)=0,\,j_\mu(\tau')=-k\delta^D_\mu\delta(\tau’-\tau)}\,.
\end{align}
The result reads
\begin{align}\label{Dt}
	\mathbb{D}_\theta=-\frac{T}{6(4\pi T)^{D/2}}\int_{\partial M} d^{D-1}x\,\sqrt{h}\,\text{tr}\,(L)\,.
\end{align}
Collecting \eqref{Db}, \eqref{Dd}, \eqref{Dt} we get the direct contributions to the first Seeley-DeWitt coefficients
\begin{align}
		a_0^{\text{dir}}&=\frac{1}{(4\pi)^{D/2}}\int_M d^Dx\sqrt{g}\,\text{tr}\,(1)
		=\frac{\text{Vol}(M)}{(4\pi)^{D/2}}\ {\text{tr}\,(1)}\,,\\[2mm]
		a_1^{\text{dir}}&=0\,,\\[2mm]
		a_2^{\text{dir}}&=\frac{1}{(4\pi )^{D/2}}\int_M d^Dx\sqrt{g}\;\text{tr}\left( \frac{R}{6}-C+\partial_\sigma \omega^\sigma+  \omega^\sigma \partial_\sigma \text{log}\sqrt{ g}-\omega^\sigma\omega_\sigma\right)+\mbox{}\nonumber\\[2mm]
		&\mbox{}+\frac{1}{(4\pi )^{D/2}}\int_{\partial M} d^{D-1}x\sqrt{h}\;\text{tr}\left( \frac{L}{3}-S+\sqrt{g_{DD}}\,\omega^D\right)\,.
\end{align}

\subsection{Indirect term}

We now compute the contribution of the second term in \eqref{extension} --correspondingly, in \eqref{pirep+intrep}-- which we call {\it indirect} contribution,
\begin{align}
	\mathbb{I}=\int_M d^Dx\sqrt{g(x)}\;\text{tr}\left(\chi\langle \tilde x| e^{-T{\mathcal D}_S}|x\rangle\right)\,.
\end{align} 
For this term ${\Delta} x^\mu=\tilde x^\mu-x^\mu=-2x^D\delta^\mu_D$, so \eqref{expr mean vaue} gives
\begin{align}\label{ind}
	\mathbb{I}=\frac{1}{(4\pi T)^{D/2}}\int_M d^Dx\,\sqrt{g}\;
	e^{-g_{DD}\frac{(x^D)^2}{2T}}
	\ \text{tr}\left(\chi\left\langle\mathcal{P} e^{-\int_0^1d\tau H_{\text{int}}(h(\tau),p(\tau))}\right\rangle\right)\,,
\end{align}
where $H_{\text{int}}$ can again be written as
\begin{align}
		H_{\text{int}}(h,p)&=H_b(h,p)- H_\theta(h,p)\,\Theta(-(1-2\tau)x^D-\sqrt{T}h^D)+\mbox{}\nonumber\\[2mm]
		&+ H_\delta(h)\delta((1-2\tau)x^D+\sqrt{T}h^D)\,.
\end{align}
Here $H_b$, $H_\theta$ and $H_\delta$ are the same as in the case of the direct contribution --eqs.\ \eqref{hb},\eqref{ht},\eqref{hd}-- but after making the replacements $x^D+\sqrt{T}h^D(\tau)\rightarrow (1-2\tau)x^D+\sqrt{T}h^D(\tau)$ and $p_\mu(\tau)\rightarrow p_\mu(\tau)-ix^D\delta^D_\mu g_{DD}(x)$. The Gaussian factor in \eqref{ind} suggests the rescaling $x^D\rightarrow \sqrt{T}x^D$. The rest of the computation proceeds along the same lines of the direct case so we simply state the final result for the indirect contributions to the Seeley-DeWitt coefficients,
\begin{align}
		a_0^{\text{ind}}&=0\,,\\[2mm]
		a_1^{\text{ind}}&=\frac{1}{(4\pi )^{D/2}}\int_{\partial M} d^{D-1}x\,\sqrt{h}\;\text{tr}\,\chi\,,\\[2mm]
		a_2^{\text{ind}}&=-\frac{1}{(4\pi )^{D/2}}\int_{\partial M} d^{D-1}x\,\sqrt{h}
		\;\text{tr}\left(S-\sqrt{g_{DD}}\,\chi\,\omega^D\right)\,.
\end{align}

\subsection{Collected contributions}

The sum of both direct and indirect contributions gives
\begin{align}
		a_0&=\frac{\text{Vol}(M)}{(4\pi)^{D/2}}\ \text{tr}(1)\,,\\[2mm]
		a_1&=\frac{1}{(4\pi )^{D/2}}\int_{\partial M} d^{D-1}x\,\sqrt{h}\;\text{tr}\,\chi\,,\\[2mm]
		a_2&=\frac{1}{(4\pi )^{D/2}}\int_M d^Dx\,\sqrt{g}\;\text{tr}\left( \frac{R}{6}-C+\partial_\mu \omega^\mu+  \omega^\mu \partial_\mu \text{log}\sqrt{ g}-\omega^\mu\omega_\mu\right)+\mbox{}\nonumber\\[2mm]
		&\mbox{}+\frac{1}{(4\pi )^{D/2}}\int_{\partial M} d^{D-1}x\,\sqrt{h}\;\text{tr}\left( \frac{L}{3}-2S+2\sqrt{g_{DD}}\ \Pi^+\omega^D\right)\,.
\end{align}
These expressions coincide with the coefficients for an operator of the form \eqref{partial deriv} reported in \cite{Vassilevich:2003xt}. By replacing $C$, $\omega^\mu$, $S$ and $\chi$ for those of the gauge operator we get
\begin{align}
		a_0(\mathcal{D})&=\frac{N\,D}{(4\pi)^{D/2}}\ \text{Vol}(M)\,,\\[2mm]
		a_1(\mathcal{D})&=\pm\frac{N(D-2)}{(4\pi )^{D/2}}\ \text{Vol}(\partial M)\,,\\[2mm]
		a_2(\mathcal{D})&=\frac{N(D-6)}{6(4\pi )^{D/2}}\int_M d^Dx\,\sqrt{g}\; R
		+\frac{N(D-6)}{3(4\pi )^{D/2}}\int_{\partial M} d^{D-1}x\,\sqrt{h}\;L\,.
\end{align}
If we replace instead $C_{gh}$, $\omega^\mu_{gh}$, $S_{gh}$ and $\chi_{gh}$ we get the coefficients for the ghost operator,
\begin{align}
		a_0(\mathcal{B})&=\frac{N}{(4\pi)^{D/2}}\ \text{Vol}(M)\,,\\[2mm]
		a_1(\mathcal{B})&=\pm\frac{N}{(4\pi )^{D/2}}\ \text{Vol}(\partial M)\,,\\[2mm]
		a_2(\mathcal{B})&=\frac{N}{6(4\pi )^{D/2}}\int_M d^Dx\,\sqrt{g}\; R
		+\frac{N}{3(4\pi )^{D/2}}\int_{\partial M} d^{D-1}x\,\sqrt{h}\;L\,.
\end{align}
Both for $\mathcal D$ and $\mathcal B$ the upper (lower) sign in $a_1$ corresponds to absolute (relative) boundary conditions. 

Finally, as shown by \eqref{eff act 2}, the UV behavior of the Yang-Mills theory is given by both the gauge and the ghost contributions as $a_n=a_n(\mathcal{D})-2a_n(\mathcal{B})$. The first of them are
\begin{align}
		a_0&=\frac{N(D-2)}{(4\pi)^{D/2}}\ \text{Vol}(M)\,,\\[2mm]
		a_1&=\pm\frac{N(D-4)}{(4\pi )^{D/2}}\ \text{Vol}(\partial M)\,,\\[2mm]
		a_2&=\frac{N(D-8)}{6(4\pi )^{D/2}}\left(\int_M d^Dx\,\sqrt{g}\; R
		+2\int_{\partial M} d^{D-1}x\,\sqrt{h}\;L\right)\,.
\end{align}

\section{Constant background field}\label{sec cbf}

In this section we turn to a different application of worldline representations, namely, the rate of gluon production due to a chromoelectric field background. {In the absence of boundaries the rate of gluon production is already known \cite{Nayak:2005yv}. Since the produced particles are massless, the result does not contain the typical non-perturbative exponential which characterizes tunneling through the mass gap; rather, the one-loop production rate follows a power-law behavior in the external field\footnote{{Higher-loop analyses in QED \cite{Affleck:1981bma,Lebedev:1984mei,Huet:2010nt,Huet:2017ydx,Huet:2018ksz,Huet:2020awq} indicate that the rate of electron/positron production might be analytic in the coupling constant $\alpha$ --a remarkable result. However, these studies focus on the weak-field limit $eE\ll m^2$ so they cannot be straightforwardly extended to the massless case.}}. Here, we consider a homogeneous background field in three-dimensional half-space. In the bulk of the space we recognize the production rate mentioned above but we also find an additional contribution located close to the boundary. At the end of the section we give an interpretation of this contribution in terms of bouncing instantons.}

We take  Euclidean 4-dimensional spacetime $M=\mathbb{R}^3\times\mathbb{R}^+$ with coordinates $x_\mu$ ($\mu=0,1,2,3$), such that $x_3\geq 0$, and introduce an homogeneous background field $E^I_i=F^I_{0i}=E\,\delta^{I1}\delta_{i2}$ (with $E$ some real constant) in some internal direction $I=1$ of the gauge group, and tangentially oriented with respect to the boundary: the boundary is $x_3=0$ and the chromoelectric field points in the $x_2$-direction\footnote{This is a strong assumption which we adopt for simplicity. Some of the present authors and collaborators have analyzed with more standard tools the case of an electric field normal to the boundary, which is technically much more complicated; the results will be presented elsewhere.}.

For this background we choose the gauge field $A^I_\mu=-E\,\delta^{I1}\delta_{\mu 0}\,x_2$,  which satisfies absolute boundary conditions at $x_3=0$. The operator $\mathcal D$ can thus be written as (see \eqref{H}) 
\begin{align}
	{\mathcal{D}=-\vec \nabla^2-(\partial_0-Ex_2\,\mathcal{F})^2-2E\, \mathcal{F}\,\mathcal{H}}\,,
\end{align}
where $\mathcal{F}$ and $\mathcal H$ are constant antisymmetric matrices which act on gauge and Lorentz indices, with elements $(\mathcal{F})^{IJ}=f^{IJ1}$ and $(\mathcal{H})^{\;\;\mu}_\nu=(\delta^\mu_0\delta^2_\nu-\delta^\mu_2\delta^0_\nu)$. According to \eqref{genbc}, absolute boundary conditions correspond to $S=0$ and $\chi=\text{diag}\,(1,1,1,-1)$. If we choose the constant extension of $\chi$ to the whole $\mathbb{R}^4$ one can easily check that $\mathcal{D}_S=\mathcal{D}$. One also finds that the operator is already Weyl ordered.

We compute the 1-loop effective action through \eqref{eff act 2} and \eqref{pirep+intrep}. The trace of the first term in \eqref{pirep+intrep} --the {\it direct} contribution-- is then
\begin{align}\label{quaact}
	&\text{tr}\,\langle x|e^{-T\mathcal D}|x\rangle
	=\int\mathcal{D}q(t)\,\mathcal{D}p(t)\ \text{tr}\left(\mathcal{P}e^{-\int_0^Tdt\left\{\vec {p}^2+
		(p_0+iE q_2\mathcal{F})^2-2E \mathcal{F}\mathcal{H}-ip_\mu\dot q_\mu\right\}}\right)\nonumber\\[2mm]
	&=\sum_i \text{tr}\left(e^{2TE \lambda_i\mathcal{H}}\right)
	\int\mathcal{D}q(t)\,\mathcal{D}p(t)\ e^{-\int_0^Tdt\left\{\vec {p}^2+
		(p_0+i \lambda_iE q_2)^2-ip_\mu\dot q_\mu\right\}}\,,
\end{align}
where $q(0)=q(T)=x$. In the second line we have introduced a sum over the eigenvalues $\lambda_i$ of $\mathcal F$ --the remaining trace then runs only over Lorentz indices. For each value of $i$ the path integral is the 4-dimensional quantum mechanical transition  amplitude of a particle of mass $m=1/2$ with initial and final points at $x$ in Euclidean time $T$ under a homogeneous magnetic field $eB=i\lambda_i E $. The result of integrating this quadratic action is well known to be \cite{Feynman Hibbs}
\begin{align}\label{vector part trace}
	\text{tr}\,\langle x|e^{-T\mathcal D}|x\rangle=\frac{1}{(4\pi T)^2}\sum_i \text{tr}\left(e^{2TE \lambda_i\mathcal{H}}\right)\frac{\lambda_i E T}{\sin(\lambda_i E T)}\,.
\end{align}

As for the second term in \eqref{pirep+intrep} --the {\it indirect} contribution-- we note that $\chi \mathcal{H}=\mathcal{H}$, so the only difference with the direct contribution are the endpoints of the trajectories, which are now $q(0)=x$ and $q(T)=\tilde x$. {Following the redefinitions and changes of variables used at the beginning of section \ref{sec ht} one obtains after some algebra}
\begin{align}
		\langle \tilde x|e^{-T\mathcal D}|x\rangle=e^{-\frac{x_3^2}{T}}\,\langle  x|e^{-T\mathcal D}|x\rangle\,.
	\end{align}
Collecting both results we conclude
\begin{align}\label{dir+indir_gauge}
	&\text{Tr}\,e^{-T\mathcal D}
	=\int d^4x\ \text{tr}\,\langle  x|e^{-T\mathcal D_S}|x\rangle+
	\int d^4x\ \text{tr}\,\chi\,\langle\tilde   x|e^{-T\mathcal D_S}|x\rangle\nonumber\\[2mm]
	&=\frac{\mathcal T\times{\rm Vol}(\partial M)}{(4\pi T)^2}
	\int_0^\infty dx_3 \left(1+e^{-\frac{x_3^2}{T}}\right)\sum_i \text{tr}\left(e^{2TE \lambda_i\mathcal{H}}\right)\frac{\lambda_i E T}{\sin(\lambda_i E T)}\,,
\end{align}
where $\mathcal T$ represents the (infinite) length of the time interval and ${\rm Vol}(\partial M)$ the (infinite) area of the boundary.

As for the ghosts fluctuation operator $\mathcal B$, absolute boundary conditions imply $\chi=1$ so, once more, direct and indirect contributions only differ in the endpoints of the worldlines. The trace can be read directly from \eqref{dir+indir_gauge} by simply omitting the factor involving the matrix $\mathcal H$ (for it acts on Lorentz indices),
\begin{align}
	\text{Tr}\,e^{-T\mathcal B}
	=\frac{\mathcal T\times{\rm Vol}(\partial M)}{(4\pi T)^2}
	\int_0^\infty dx_3 \left(1+e^{-\frac{x_3^2}{T}}\right)\sum_i \frac{\lambda_i E T}{\sin(\lambda_i E T)}\,.
\end{align}
Collecting all results and using $\mathcal{H}^3=-\mathcal{H}$ to compute the trace, the 1-loop effective action reads
\begin{align}
		\Gamma[A]&=-\frac{\mathcal T\,{\rm Vol}(\partial M)}{16\pi^2}
		\int_0^\infty\frac{dT}{T^2} \int_0^\infty dx_3
		\left(1+e^{-\frac{x_3^2}{T}}\right)
		\sum_i \frac{\cos(2\lambda_i E T)}{\sin(\lambda_i E T)}\,\lambda_i E\,.
\end{align}
The rate of gluon production is given by twice the {imaginary} part of the {Minkowskian} effective action $\Gamma_M[A]$. {By definition, $\Gamma[A]=-i\Gamma_M[A]$ once we undo the Wick rotation. The Euclidean action is the result (up to the just mentioned overall $-i$ factor) of analytically extending the Minkowskian action to the positive imaginary half-axis in the complex variable $A_0$ and to the negative imaginary half-axis in the time coordinate. To undo the rotation we then replace $\mathcal T\to i \mathcal T$ and $E\to -iE$ and take the limit where the new $\mathcal T,E$ lay on the real axis. For the electric field it is necessary to make explicit that we are taking the limit form the upper complex half-space, so we write $E\to -i(E+i0)$}.

For the special unitary groups the antisymmetric matrix $\mathcal{F}$ is also real so its eigenvalues are purely imaginary conjugate pairs, $\lambda_i=\pm i\alpha_i$, with $\alpha_i\in \mathbb{R}^+$ (note that zero eigenvalues do not contribute to the {imaginary} part of $\Gamma_M[A]$). In terms of the Minkowskian action we thus obtain
\begin{align}
		\frac{2\,\text{Im}\,\Gamma_M[A]}{\mathcal T\,{\rm Vol}(\partial M)}
		&=\sum_i\frac{\alpha_i E}{4\pi^2}\ {{\rm Im}}\int_0^\infty dx_3\int_0^\infty
		\frac{dT}{T^2}\left(1+e^{-\frac{x_3^2}{T}}\right)
		\frac{\text{cos}(2\alpha_i E T)}{\text{sin}(\alpha_i (E+i0) T)}\,.
\end{align}
Finally, contributions to the imaginary part stem from the singularities at $\alpha_i |E| T =\pi n$, with $n=1,2,3,\dots$,
\begin{align}\label{last}
	\frac{2\,\text{Im}\,\Gamma_M[A]}{\mathcal T\times{\rm Vol}(\partial M)}
	&=\sum_i\frac{1}{4\pi}\int_0^\infty dx_3\int_0^\infty
	\frac{dT}{T^2}\left(1+e^{-\frac{x_3^2}{T}}\right)
	\sum_{n=1}^\infty (-1)^{n+1}\delta(T-\tfrac{\pi n}{\alpha_i|E|})\nonumber\\[2mm]
	&=\int_0^\infty dx_3	\sum_i
	\alpha_i^2\left[\frac{|E|^2}{48\pi}+
	\frac{|E|^2}{4\pi^3}\,
	\sum_{n=1}^\infty\frac{(-1)^{n+1}}{n^2}\,e^{-\frac{\alpha_i|E|}{\pi n}x_3^2}\right]\nonumber\\[2mm]
	&=	\sum_i
	\left[\frac{(\alpha_i|E|)^2}{48\pi}\,L+
	(1-\tfrac{1}{\sqrt2})\,\zeta(\tfrac32)\,
	\frac{(\alpha_i|E|)^\frac32}{8\pi^2}\right]
	 \,.
\end{align}
The total length in the normal direction to the boundary is represented by $L$; $\zeta$ is the Riemann $\zeta$-function. We see that apart from the bulk rate of gluon production \cite{Nayak:2005yv} --proportional to the volume-- there is an additional boundary contribution --proportional to its area-- which occurs in a thin layer of width $\sim 1/\sqrt{|E|}$ {near} the boundary (see the second line in \eqref{last}).

For the specific case of QCD, the structure constants of $su(3)$ give the values $\alpha_1=1$, $\alpha_2=\alpha_3=\frac12$ so the rate of gluon production is
\begin{align}
	\frac{2\,\text{Im}\,\Gamma_M[A]}{\mathcal T}
	={\rm Vol}(M)\ \frac{1}{32\pi}\,|E|^2+
	{\rm Vol}(\partial M)\ \frac{\zeta(\tfrac32)}{16\pi^2}\,|E|^\frac32
	\,.
\end{align}

We conclude with an important remark. Note that the path integral \eqref{quaact} --being quadratic in the phase-space coordinates-- can be integrated exactly, giving \eqref{vector part trace}. Alternatively, one could use saddle-point approximation around classical trajectories --{\it worldline instantons}--, as originally done in \cite{Affleck:1981bma}. In this seminal article the classical trajectories are circles and their actions eventually give the usual Schwinger factors $e^{-\pi m^2 n/eE}$, where $E$ is the external electric field, $e$ and $m$ the electron's charge and mass, and $n$ represents the winding number of the classical solution. In our example, since the gluons are massless, such exponential factors are absent. Nevertheless, the presence of a boundary allows the existence of helical trajectories which are closed due to a bounce at $x_3=0$.

\begin{figure}
	\centering
	\begin{minipage}{.8\linewidth}
		\includegraphics[width=10cm]{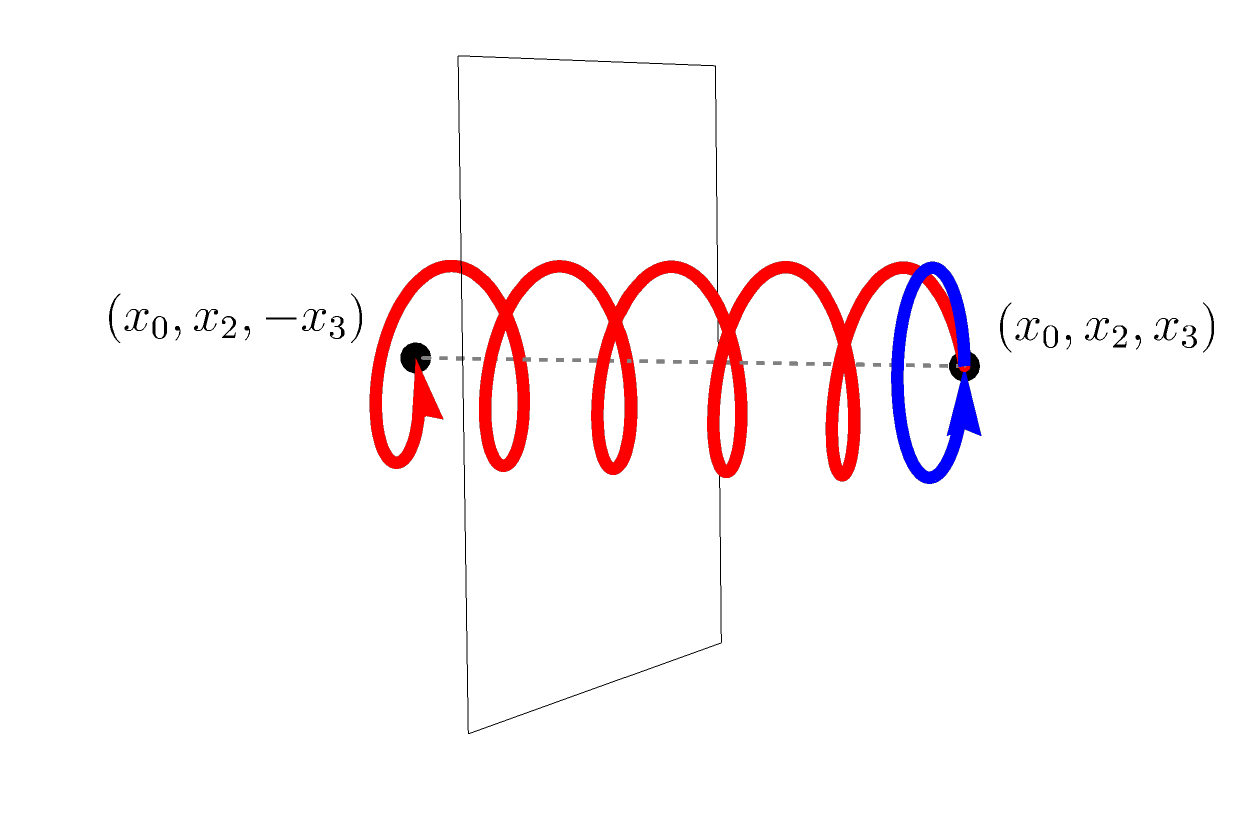}
		\caption{\small Instantons with $n=1$ corresponding to a direct (blue) contribution and with $n=5$ corresponding to an indirect (red) contribution. The positive integer $n$ corresponds to the index $n$ in \eqref{last} that refers to the singularities in the heat trace.}
		\label{instantons}
	\end{minipage}\hspace{10mm}
\end{figure}

In fact, a classical solution of the action given in \eqref{quaact} but with antiperiodic boundary conditions in the coordinate $q_3(t)$ is given by
\begin{align}
	q_3(t)=\left(1-\frac {2t}T\right)x_3
\end{align}
together with a circular motion in the plane $q_0$-$q_2$ with arbitrary radius and frequency $|2\lambda_i E|$. This is represented in figure \ref{instantons} by the helix ending at the image point across the boundary, with winding number $n=5$.
Due to the translation in the $x_3$ direction the action is not vanishing but takes the value
\begin{align}
	S[q(t),p(t)]=\frac{x_3^2}{T}=\frac{|\lambda_i E|}{\pi n}\,x_3^2\,.
\end{align}
We have used $T=\pi n/|\lambda_i E|$, which is imposed by periodicity of the circular motion. Upon the replacements $E\to -iE$ and $\lambda_i\to\pm i\alpha_i$ we reproduce the exponential factor in the boundary contribution of the second line of \eqref{last}. We find it interesting that boundary contributions can be read from worldline instantons that bounce at the boundary or, equivalently, joins an arbitrary point with its image across the boundary. Note that the use of worldline instantons allows one to explore non-quadratic actions.

\section{Conclusions}\label{sec conclu}

In this work we developed a worldline description for the heat kernel of the quantum fluctuation operator associated to a Yang-Mills theory  in the presence of a boundary, background fields and curvature. We considered the case of the $D$-dimensional manifold $M=\mathbb{R}^{D-1}\times \mathbb{R}^+$, the boundary being at $x^D=0$ and the metric fulfilling $g_{\alpha D}=0$ for $\alpha\neq D$ and studied two kinds of mixed boundary conditions called relative and absolute conditions \cite{Vassilevich:2003xt} (see section 2). We did this in section 3 following the work done in \cite{Corradini:2019nbb} for scalars and in \cite{Manzo:2024gto} for fermions, that is, by properly extending every relevant quantity defined on $M$ to an extended manifold $\tilde M=\mathbb{R}^D$, which has no boundary, and solving the heat equation via method of images. Equation \eqref{pirep+intrep} is the result of this procedure and the centerpiece of this article. Since the heat kernel is directly related to the one-loop effective action of the theory, this expression has many applications.

In section 4 we used it to compute the first three Seeley-DeWitt coefficients, which contain the structure of the leading UV divergences of the theory at one-loop order. These are in coincidence with those obtained in \cite{Vassilevich:2003xt} and thus provide a check of our formula.

In the last section we used the representation \eqref{pirep+intrep} of the quantum transition amplitude to compute the imaginary part of the effective action for Yang-Mills theory in the presence of a boundary and a constant chromoelectric background under absolute boundary conditions. According to \eqref{pirep+intrep}, two different types of contributions --dubbed direct and indirect-- arise. The result can be interpreted in terms of the classical solutions of the path integral action ({\it worldline instantons}) either in phase space or in configuration space. The direct part can be computed in terms of the well known trajectories corresponding to the circular motion of a charged particle in a homogeneous magnetic field and coincides with the known result for the case without boundaries obtained in \cite{Nayak:2005yv}. The effects of the boundary are relevant within a collar neighborhood of width $\sim |E|^{-1/2}$ and come from the indirect part. It receives contributions from trajectories which are antiperiodic in the coordinate normal to the boundary and represents an instanton that reaches the image point or, alternatively, an instanton which bounces at the boundary. As far as we know, this type of worldline instantons that appear in the presence of boundaries, had not been used in the literature. We think there is a number of scenarios worth considering where these bouncing solutions might be helpful. In particular, we are currently studying different settings of the Schwinger effect but in the more involved situation of an electric field perpendicular to the boundary.

As for other applications of our results, we remark that the 1-loop effective action --for which we give here a worldline representation-- also contains the information on anomalies, $N$-point functions, etc. {The  procedure presented in this article could also be applied in the context of open worldlines, which are used to compute the complete propagator in the presence of a background. Note however that the derivation of master formulas for gluon amplitudes --one of the most useful applications of the worldline framework-- strongly relies in the quantization of the spinning particle \cite{Corradini:2015tik,Bastianelli:2025xne}. Indeed, in the bottom-up approach the auxiliary Grassmann fields are used not only to represent the Hilbert space of spin and color degrees of freedom but also to build the appropriate vertex operators. This is, a priori, a shortcoming of our approach which motivates the search of an implementation of boundaries in the spinning particle formulation.}

To conclude, we give a word on what future work could entail. {We have already checked that our technique based on the method of images works well for two flat boundaries facing each other under different combinations of absolute and relative boundary conditions. 
Apart from the aforementioned use of bouncing instantons for more convoluted scenarios with one single boundary, we think that they could be particularly useful to analyze other configurations of disconnected boundaries.}

\vspace{1cm}

\noindent{\bf Acknowledgments:} We thank support from CONICET (PIP 0262), UNLP (I+D X909) and DAAD (Scientific Literature Programme). LM also acknowledges support from Departamento de Física (Programa de Retención de Recursos Humanos).



\end{document}